\documentclass[%twocolumn,
showpacs,preprintnumbers,amsmath,amssymb]{revtex4}
\usepackage{amsmath}
\usepackage{empheq}
\usepackage{graphicx}
\begin{document}

\title{DISSIPATIVE UNIVERSE-INFLATION WITH SOFT SINGULARITY}

\author{   I. Brevik$^{1}$\footnote{E-mail:iver.h.brevik@ntnu.no} and A. V. Timoshkin$^{2}${\footnote{E-mail:alex.timosh@rambler.ru}}}

\medskip

\affiliation{$^{1}$Department of Energy and Process Engineering, Norwegian University
of Science and Technology, N-7491 Trondheim, Norway}

\affiliation{$^{2}$Tomsk State Pedagogical University, Kievskaja Street 60, 634050 Tomsk, Russia;
Tomsk State University  of  Control Systems and Radioelectronics, Lenin Avenue 40, 634050 Tomsk, Russia}

 \today

\begin{abstract}
We investigate the early-time accelerated universe after the Big Bang. We pay attention to the dissipative properties of the inflationary universe in the presence of a soft type singularity, making use of the parameters of the generalized equation of state of the fluid. Flat Friedmann-Robertson-Walker metric is being used. We consider cosmological models leading to the so-called type IV singular inflation. Our obtained theoretical results are compared with observational data from the Planck satellite. The theoretical predictions for the spectral index turn out to be in agreement with the data, while for the scalar-to tensor ratio there are minor deviations.

\end{abstract}

 \maketitle
\section{Introduction}

The discovery of an accelerating expansion of the late-time universe at the end of the last century \cite{riess98,perlmutter99} led to the appearance of various cosmological models, including the Big Rip \cite{caldwell03,nojiri03}, the Little Rip \cite{frampton11,brevik11,frampton12}, the Pseudo-Rip \cite{frampton12a}, the Quasi-Rip \cite{wei12} and the bounce cosmology \cite{brandenberger11,novello08,bamba14,cai11}.
After the Big Bang an early-time acceleration period is believed to exist \cite{linde08,gorbunov11}, in which the total energy and the scale  factor becomes exponentially large \cite{brandenberger10}. The inflationary era can be modeled  in terms of a cosmic fluid satisfying an inhomogeneous equation of state \cite{nojiri11}, and can be considered as a modified theory of gravity \cite{capozziello06}.

We intend to study different forms of the Hubble parameter that can reproduce the evolution of the inflationary universe in the presence of a so-called type IV finite time singularities. The first classification of finite time singularities was given in Ref.~\cite{nojiri05}. The type IV singularity is characterized by physical quantities such as the scale factor, the effective energy density, and the effective pressure, which tend to finite values near the time of the singularity, whereas the higher derivatives of the Hubble parameter diverge. This is thus a soft type of singularity. According to these properties, the  development of the universe proceeds smoothly without crushing after passage through the singularity.

Our motivation for studying this kind of singularity is related to the circumstance that this type of cosmological evolution occurs in the inflationary era \cite{nojiri15}. The  scenario with the appearance of a type IV singularity at the end of the inflation was actually predicted in Refs.~ \cite{21,nojiri05a,nojiri15a,odintsov15}.

The present  paper investigates the influence from heat dissipation on the thermodynamic parameters in the modified equation of state, in the inflationary region. We assume a flat, homogeneous and isotropic, Friedmann-Robertson-Walker spacetime. We will describe the energy dissipation via an inhomogeneous equation of state, similarly as we did in a recent paper \cite{21}. This is equivalent physically to the introduction of a bulk viscosity (the shear viscosity being omitted due to spatial isotropy). Inclusion of  a bulk viscosity in the modified equation of state may be important in the exploration of  all kinds of singularities \cite{odintsov15,brevik06,nojiri05b,brevik10,brevik14}. The dissipative universe may also be described in terms of an entropic cosmology model \cite{komatsu14} (see, for example, Ref.~\cite{brevik15}).

We moreover calculate for each of the models the inflationary parameters, the spectral index, and the tensor-to-scalar ratio, and  discuss how these theoretical results compare with the Planck experimental data.

\section{	Fluid description of a dissipative singular inflation}

In this section we will investigate the imperfect fluid in standard Einstein-Hilbert theory of gravity. The Friedmann equation for the Hubble rate is
\begin{equation}
\rho=\frac{3}{k^2}H^2, \label{1}
\end{equation}
where $\rho$ is the energy density, $H(t)=\dot{a}(t)/a(t)$ the Hubble parameter, $a(t)$ the scale factor, and $k^2=8\pi G$ with $G$    being Newton's gravitational constant. A  dot denotes derivative with respect to cosmic time $t$.

We consider a flat Friedmann-Robertson-Walker metric
\begin{equation}
ds^2=-dt^2+a^2(t)\sum_idx_i^2. \label{2}
\end{equation}
We will describe the inflationary universe as a fluid that  obeys a linear inhomogeneous equation of state,
\begin{equation}
p=\omega(t)\rho+aH^\alpha +b \dot{H}^\beta, \label{3}
\end{equation}
where $p$ is the pressure of the fluid, the thermodynamic parameter $\omega(t)$ being dependent on time $t$. Further, $a,b$ are positive dimensional constants, and the parameters $\alpha, \beta$ are positive. The second and third terms in the phenomenological equation of state (\ref{3}) describe the influence from dissipation in the inflationary universe.

We suppose that the parameter $\omega(t)$ depends linearly on time \cite{brevik07}, i.e.
\begin{equation}
\omega(t)=ct+d, \label{4}
\end{equation}
where $c,d$ are arbitrary constants.

We write the energy conservation  equation as
\begin{equation}
\dot{\rho}+3H(\rho+p)=0. \label{5}
\end{equation}

 It may be surprising that this equation has the same form as for a conventional {\it nonviscous} fluid. Usually, on admits a term containing the bulk viscosity on the right hand side when dealing with viscous cosmology. The point here is however that we take into account the properties from viscosity through the inhomogeneous equation state, instead of through the standard bulk viscosity term.

 Taking into account Eqs.~(\ref{1}), (\ref{3}) and (\ref{4}), Eq.~(\ref{5}) takes the form
 \begin{equation}
 \frac{2}{k^2}\dot{H}+\frac{3}{k^2}(ct+d+1)H^2+aH^\alpha +b {\dot{H}}^\beta=0. \label{6}
 \end{equation}
 Let us assume the parameter values $\alpha=2, \beta=1$ in this equation of motion. For the constants $a,b,c,d$ we obtain
 \begin{equation}
\left\{ \begin{array}{llll}
a=-\frac{3}{k^2}( d+1) \\
b= -\frac{2}{k^2} \\
c=0 \\
d \in{R}.
\end{array}
\right. \label{7}
\end{equation}

We now consider some singular inflation models where the  Hubble parameter is taken to have  forms that can reproduce type II and type IV singularities \cite{nojiri15},
\begin{equation}
\noindent {\rm{A.}}  \quad H(t)=f_1(t)+f_2(t)(t_s-t)^\gamma, \label{8}
\end{equation}
where $t_s$  is the time when the singularity appears. In this model the functions $f_1(t)$  and $f_2(t)$  are smooth and differentiable. If the parameter $\gamma$ satisfies $0<\gamma <1$, this form for the Hubble parameter  leads to the type II singularity. In order to get a type IV singularity one must have $\gamma >1$, where $\gamma \notin{Z}$.

For simplicity we choose both functions $f_1(t)$   and $f_2(t)$ to be  arbitrary positive dimensional constants: $f_1(t)=f_0$ and  $f_2(t)=g_0$.  This choice  is motivated by the very complicated form the equation of state would have if these functions were allowed to have an arbitrary analytic form.

In particular, if $\gamma =3/2$ in Eq.~(\ref{2}) we obtain the following simple form for the Hubble parameter,
\begin{equation}
H(t)=f_0+g_0(t-t_s)^{3/2}, \label{9}
\end{equation}
which describes type IV singularity in the inflation.

The corresponding modified equation of state gets the following form,
\begin{equation}
p=-d\rho -\frac{3}{k^2}\left\{ (d+1)\left[f_0+g_0(t-t_s)^{3/2}\right]^2 -g_0(t_s-t)^{1/2}\right\}, \label{10}
\end{equation}
which describes a type IV inflation. The second term in this equation modifies the effective pressure and describes the influence from dissipation.

Another interesting inflationary model is represented by \cite{nojiri15}
\begin{equation}
\noindent {\rm B.} \quad H(t)=h_0\left[ \left(\frac{t-t_0}{t_1}\right)^{-2n}+1\right]^{-\frac{\gamma}{2n}}, \label{11}
\end{equation}
where $h_0, t_0,t_1,\gamma$ are constants. For physical reasons it follows that $h_0>0, n>0$ and $\gamma>0$. We see that this model produces a type IV singularity at $t \approx t_0$, because
the Hubble parameter is equivalent to $H \approx h_0\left( \frac{t-t_0}{t_1}\right)^\gamma.$

We will investigate the particular case $\gamma=2n, n=3/4$.   The Hubble parameter can then be simplified so that in the vicinity of $t_0$,
\begin{equation}
H(t)=h_0\left( \frac{t-t_0}{t_1}\right)^{3/2}. \label{12}
\end{equation}
In this case the modified equation of state can be written in the form
\begin{equation}
p=d\rho -\frac{3}{k^2}(d+1)h_0^2\left( \frac{t-t_0}{t_1}\right)^3 -\frac{2}{k^2}\frac{h_0}{t_1^{3/2}}(t-t_0)^{1/2}. \label{13}
\end{equation}
Let us choose parameter values $\alpha=\beta=1$ in Eq.~(\ref{3}). Then we obtain the approximate form of equation of state,
\begin{equation}
p=-\rho-\frac{2}{k^2}\frac{h_0}{t_1^{3/2}}(t-t_0)^{1/2}. \label{14}
\end{equation}
This form coincides with Eq.~(\ref{13}) when $d=-1$.

Now, let us study the case in which the Hubble parameter has the following form \cite{nojiri15},
\begin{equation}
\noindent {\rm C.} \quad H(t)=f_0(t-t_1)^\gamma +c_0(t-t_2)^\delta, \label{15}
\end{equation}
where $f_0$  and $c_0$   are constant positive parameters, and  $\gamma, \delta >1$, but $\gamma, \delta \notin Z$. There are two type IV singularities, occurring at $t=t_1$ and $t=t_2$,  in this model of  inflation. We assume that  $t_1$ is at the end of the inflationary era and $t_2$   is any late time (may be even the present time). We consider the case $\gamma=\delta=3/2$   and take the values $\alpha=2, \beta=1$   in the equation of state (\ref{3}). The values of the parameters given in Eq.~(\ref{14}) satisfy this equation.

Analogously to the previous case, the modified equation of state can be written as
\begin{equation}
p=d\rho-\frac{3}{k^2}(d+1)[f_0(t-t_1)^{3/2}+c_0(t-t_2)^{3/2}]^2-\frac{3}{k^2}[f_0(t-t_1)^{1/2}+c_0(t-t_2)^{1/2}]. \label{16}
\end{equation}
This equation of state describes the dissipative properties of the inflationary universe.

Next, let us investigate another interesting case in which the Hubble parameter has the following form \cite{nojiri15},
\begin{equation}
\noindent {\rm D.} \quad H(t)=\frac{f_1}{\sqrt{t^2+t_0^2}}+\frac{f_2t^2(-t+t_1)^\gamma}{t^4+t_0^4}+f_3(-t+t_2)^\delta. \label{17}
\end{equation}
In order to fulfil the condition $H(t)>0$, the constant parameters $\alpha, \beta, t_0, f_1, f_2, f_3$ must be positive. Let us assume, as above, that $t_1$ corresponds to early-time universe evolution and that $t_2$ corresponds to late time. The most interesting scenario takes place if $\gamma >1, \delta >1$,  where $\gamma, \delta \notin Z$, since then a type IV singularity occurs at both early time and late time.

We will now obtain analytic approximations for the equation of state in the vicinity of each of the singularities.

Near the early-time singularity $t \approx t_1$, choosing parameter values $\gamma=\delta=3/2$, we obtain the following energy conservation equation of state,
\begin{equation}
p=d\rho-\frac{3}{k^2}(d+1)\left[ \frac{f_1}{(t^2+t_0^2)^{1/2}}+f_3(-t+t_2)^{3/2}\right]^2+\frac{2}{k^2}\left[ \frac{f_1t}{(t^2+t_0^2)^{3/2}}+\frac{3}{2}f_3(-t+t_2)^{1/2}\right]. \label{18}
\end{equation}
Thus, the late-time inhomogeneous equation of state defines the behavior of the inflation near the early-type IV singularity.

Proceeding to the late-time singularity at  $t \approx t_2$  the approximate equation of state becomes
\begin{align}
p=d\rho-\frac{3}{k^2}(d+1)\left[ \frac{f_1}{(\sqrt{t^2+ t_0^2}}+\frac{f_2t^2(-t+t_1)^{3/2}}{t^4+t_0^4}\right]^2     \\ \notag
-\frac{2}{k^2}\left[ -\frac{f_1t}{(t^2+t_0^2)^{3/2}}+f_2t\frac{(-t+t_1)^{1/2}}{(t^4+t_0^4)^2}\left( \frac{7}{2}t^5-3t_1t^4+t_1t_0^4-\frac{1}{2}t_0^4t\right)\right]  \label{19}
\end{align}

\noindent Analogously to the example above, the early-time inhomogeneous equation of state defines the behavior of the inflation near the late-time type IV singularity.

Thus, we have described  dissipation in the inflationary universe in terms of a model having a soft type singularity, choosing appropriate values for the parameters in the generalized equation of state.

\section{	Comparison of parameters for singular dissipative inflation with recent observational data}

In the previous section we described  singular inflation with dissipation in  terms of a modified  equation of state for the fluid. Now, the presence of  type IV singularities can have an influence on the parameters of the inflation. Therefore, this section is devoted to calculation of the slow-roll parameters and also the following observational indices: the spectral index of primordial curvature perturbations,  and the scalar-to-tensor ratio. We then wish  to compare the theory with the recent Planck observational data \cite{ade14}. A detailed   analysis of the behavior of the slow-roll parameters for a perfect-fluid modified equation of state was given in Ref.~\cite{bamba14a}. We intend to give, for each  inflationary model, a qualitative analysis of how these  observational indices compare with the theoretical predictions of our models.

Let us  introduce the slow-roll parameters \cite{myrzakulov15},
\begin{equation}
\varepsilon=-\frac{\dot{H}}{H^2}, \quad \eta=\varepsilon-\frac{\dot{\varepsilon}}{2\varepsilon H}. \label{20}
\end{equation}
We can express the acceleration of the universe via the parameter $\varepsilon$,
\begin{equation}
\frac{\ddot{a}}{a}=H^2+\dot{H}=H^2(1-\varepsilon). \label{21}
\end{equation}
Acceleration of the universe corresponds to $\ddot{a}/a >0$, what implies $\varepsilon <1$.

From the slow-roll parameters we can calculate the spectral index $n_s$ and the scalar-to-tensor ratio $r$,
\begin{equation}
n_s=1-6\varepsilon +2\eta, \quad r=16\varepsilon. \label{22}
\end{equation}

Let us calculate these parameters for the model where $H(t)=f_0+g_0\Delta t_s^{3/2}$. We obtain
\begin{equation}
\varepsilon=\frac{3g_0{\Delta t_s}^{1/2}}{2(f_0+g_0\Delta {t_s}^{3/2})^2}, \quad \eta=
\frac{11 g_0\Delta {t_s}^{3/2}-f_0}{4\Delta t_s(f_0+g_0\Delta {t_s}^{3/2})^2}, \label{23}
\end{equation}
where $\Delta t_s=t_s-t.$

The spectral index $n_s$ and the scalar-to-tensor ratio $r$ become in this case
\begin{equation}
n_s-1=-\frac{7g_0\Delta {t_s}^{3/2}+f_0}{2\Delta t_s(f_0+g_0\Delta {t_s}^{3/2})^2}, \quad r=\frac{24g_0\Delta {t_s}^{1/2}}{(f_0+g_0\Delta {t_s}^{3/2})^2}. \label{24}
\end{equation}
From this we see that singularities in the inflation occur at $t=t_s-(f_0/g_0)^{2/3}$.

From observations by the Planck satellite it is known that $n_s=0.9603 \pm 0.0073$ and $r<0.11$ \cite{ade14}.

Let us consider the particular case when $f_0=0, g_0=1$.  If we require that $\Delta t_s \approx 19.81$, then the spectral index coincides with the result of the Planck measurement. However, the scalar-to-tensor ratio $r \approx 0.272$ and  there is a  deviation from the data. Consequently, the results of the Planck experiment cannot be realized completely in this model.

In another inflationary model where the Hubble parameter is $H(t)=h_0\left( \frac{t-t_0}{t_1}\right)^{3/2}$,   the slow roll parameters take the values
\begin{equation}
\varepsilon \approx \frac{3}{2t_0}\left( 1+\frac{t}{t_0}\right), \quad \eta \approx \varepsilon, \label{25}
\end{equation}
where $|t| \leq t_0.$  In this case the spectral index $n_s$  and the scalar-to-tensor ratio $r$ become
\begin{equation}
n_s-1 \approx -4\varepsilon, \quad r \approx \frac{24}{t_0}\left( 1+\frac{t}{t_0}\right). \label{26}
\end{equation}
We get concordance between the calculated spectral index and experiment if we impose the restriction $\varepsilon \approx 0.01$. However, the scalar-to-tensor ratio $r\approx 0.16$, and so there is some deviation from the observed data. Thus, the present model cannot reproduce the Planck observations.

Now we discuss the next model "D" of  inflation near the early-time singularity, when  $H(t)=\frac{f_1}{\sqrt{t^2+t_0^2}}+f_3(t-t_2)^{3/2}$. Suppose that $|t| \leq t_0$, and take into account that $|t| \approx t_1$ (the early-time evolution). A calculation of the slow-roll parameters at $f_1=0$ gives the result
\begin{equation}
\varepsilon = -\frac{3}{2t_0^3}\frac{1}{t-t_2}, \quad \eta=\varepsilon \left[1-\frac{1}{3t_0^3f_3(t-t_2)^{3/2}}\right]. \label{27}
\end{equation}
Here the singularity occurs at $t=t_2$.

 The spectral index is equal to $n_s=1-4\varepsilon -\frac{2}{3t_0^3f_3(t-t_2)^{3/2}}.$ The scalar-to-tensor ratio coincides with the result of the Planck experiment if the slow-roll parameter satisfies $\varepsilon < 0.0687$. Consequently, the spectral index satisfies the Planck bound, as when $f_3$ is positive the left hand side of the inequality  $\frac{2}{3t_0^3f_3(t-t_2)^{3/2}}>-0.2351$ becomes positive and the inequality thus trivial.  So, the results of the Planck experiment can be completely realized in this model.

Analogously,  we can consider  inflation near the late-time singularity, when $ H(t)=\frac{f_1}{\sqrt{t^2+t_0^2}} +\frac{f_2t^2(t-t_1)^{3/2}}{t^4+t_0^4}$.  If we suppose that  $|t| \approx t_2$ (the late-time evolution), and $|t| \leq t_0$, then by keeping only higher order terms in $t$ we obtain at $f_1=0$  the following approximation for the slow-roll parameters,
\begin{equation}
\varepsilon \approx -\frac{7}{2t}, \quad \eta =\varepsilon \left( 1-\frac{t_0^4}{7f_2t^{7/2}}\right).  \label{28}
\end{equation}
The spectral index is equal to $n_s=1-4\varepsilon -\frac{2t_0^4}{7f_2t^{7/2}}.$ As before, the scalar-to-tensor ratio coincides with the result of the Planck experiment if the slow-roll parameter satisfies $\varepsilon < 0.0687$. Consequently, the spectral index satisfies the Planck experiment, because when $f_2$ is positive the left hand side of the inequality  $\frac{2t_0^4}{7f_2t^{7/2}}> -0.2351$ is positive and the inequality again trivial.  The results of the Planck experiment can be realized completely also in this model.

 Finally, we discuss the model "C" near the singularity $t \approx t_1$, when the Hubble parameter equals $H(t)=c_0(t-t_2)^{3/2}.$ In this approximation we obtain
 \begin{equation}
 \varepsilon =-\frac{2}{c_0\Delta t_2^3}, \quad \eta=\varepsilon +\frac{1}{2c_0\Delta {t_2}^{3/2}}, \label{29}
 \end{equation}
 where $\Delta t_2=t-t_2$. Here the singularity occurs at $t=t_2$.

 The spectral index is equal to $n_s=1-4\varepsilon +\frac{1}{c_0\Delta {t_2}^{5/2}}$. If it fulfils the requirement $\frac{1}{c_0\Delta {t_2}^{5/2}}<0.2351$, we obtain agreement with the Planck experiment.

 The case of model "C" near the singularity $t\approx t_2$ coincides with the previous one if we repeat the calculation along the same lines.

Note that the recent BICEP2 experiments \cite{jimenez16} have fixed the cosmic radiation with a scalar-to-tensor ratio  $r=0.20^{+0.07}_{-0.05}$.
 This result confirms the usefulness of the considered models.

\section{Conclusion}

In this paper we have considered  inflation in the presence of dissipation,  in terms of  parameters in the assumed inhomogeneous equation of state for the cosmic fluid. We studied fluid models leading to the so-called  type IV singular inflation. Our method of adding  dynamic and dissipative terms to the equation of state,  corresponds to viscous fluid cosmology. We  studied type IV singular inflation models because this type of  singularity appears in the inflationary period. Recently, there have appeared   experimental data  favoring the existence of  type IV singularity in the  inflationary period \cite{jimenez16}

We   calculated central parameters of the inflation in each of the theoretical models,  and compared them with  observational data. As a result we concluded that all models considered in this section describe an acceleration of the inflationary universe. The inflationary parameters contain singular points, meaning physically the presence of  instabilities in the dynamic system. Observations from the Planck satellite show that the results for the spectral index of curvature perturbations  can be reproduced in all our studied models, but as regards  the scalar-to-tensor ratio there are small deviations from the experimental data.

\bigskip

\section*{Acknowledgements}
              This work was supported by a grant from the Russian Ministry of Education and Science, project TSPU-139 (A.V.T.).

\end{document}